\documentclass[prd,aps,twocolumn,superscriptaddress]{revtex4-2}

\usepackage[T1]{fontenc}
\usepackage[latin9]{inputenc}
\usepackage{amssymb}
\usepackage{graphicx}
\usepackage{amsmath,color}
\usepackage{mathrsfs}
\usepackage{float}
\usepackage{indentfirst}
\usepackage{subfigure}
\usepackage{multirow}
\usepackage{tabu}
\usepackage{threeparttable}
\usepackage{booktabs}
\usepackage{txfonts}
\usepackage{amsmath,amssymb,bbm}
\usepackage[normalem]{ulem}
\usepackage[euler]{textgreek}
\usepackage{bm}% bold math

\usepackage[dvipsnames]{xcolor}
\usepackage{soul}
\usepackage[colorlinks=true,urlcolor=MidnightBlue,citecolor=MidnightBlue,linkcolor=MidnightBlue]{hyperref}

\newcommand\Tstrut{\rule{0pt}{3.1ex}}         % = `top' strut
\newcommand{\tmmathbf}[1]{\boldsymbol{#1}}
\newcommand{\mathd}{\mathrm{d}}
\newcommand{\mathD}{\mathcal{D}}
\newcommand{\mathi}{\mathrm{i}}
\newcommand{\mathe}{\mathrm{e}}
\newcounter{nnnote}

{\newtheorem{note*}[nnnote]{Note}}
\newcommand{\boldn}{\tmmathbf{n}}
\newcommand{\boldl}{\tmmathbf{l}}
\newcommand{\boldm}{\tmmathbf{m}}
\newcommand{\boldL}{\tmmathbf{L}}

\newcommand{\sfQ}{\mathsf{Q}}
\newcommand{\sfL}{\boldsymbol{\mathsf{L}}}
\newcommand{\sfR}{\boldsymbol{\mathsf{R}}}
\newcommand{\sfP}{\boldsymbol{\mathsf{P}}}

\newcommand{\lmax}{l_{\mathrm{max}}}

\newcommand{\Eq}[1]{Eq.~(\ref{#1})}
\newcommand{\<}{\langle}
\renewcommand{\>}{\rangle}

\newcommand{\tr}{\mathrm{Tr}}

\newcommand\redsout{\bgroup\markoverwith{\textcolor{red}{\rule[0.5ex]{2pt}{0.4pt}}}\ULon}

\usepackage{scalerel}

\begin{document}

\preprint{APS/123-QED}

\title{Tensor network simulation of the (1+1)-dimensional \boldmath$O(3)$ nonlinear \boldmath$\sigma$-model with \boldmath$\theta=\pi$ term}%

\author{Wei Tang}
\affiliation{International Center for Quantum Materials, School of Physics, Peking University, Beijing 100871, China}

\author{X. C. Xie}
\affiliation{International Center for Quantum Materials, School of Physics, Peking University, Beijing 100871, China}

\author{Lei Wang}
\email{wanglei@iphy.ac.cn}
\affiliation{Beijing National Lab for Condensed Matter Physics and Institute of Physics, Chinese Academy of Sciences, Beijing 100190, China}
\affiliation{Songshan Lake Materials Laboratory, Dongguan, Guangdong 523808, China}

\author{Hong-Hao Tu}
\email{hong-hao.tu@tu-dresden.de}
\affiliation{Institute of Theoretical Physics, Technische Universit\"at Dresden, 01062 Dresden, Germany}

\date{\today}

\begin{abstract}
We perform a tensor network simulation of the (1+1)-dimensional $O(3)$ nonlinear $\sigma$-model with $\theta=\pi$ term.  
Within the Hamiltonian formulation, this field theory emerges as the finite-temperature partition function of a modified quantum rotor model decorated with magnetic monopoles.
Using the monopole harmonics basis, we derive the matrix representation for this modified quantum rotor model, which enables tensor network simulations.
We employ our recently developed continuous matrix product operator method [\href{https://journals.aps.org/prl/abstract/10.1103/PhysRevLett.125.170604}{Tang \emph{et al.}, Phys.~Rev.~Lett. 125, 170604 (2020)}] to study the finite-temperature properties of this model and reveal its massless nature.
The central charge as a function of the coupling constant is directly extracted in our calculations and compared with field theory predictions.
\end{abstract}

\maketitle

%\tableofcontents

\section{Introduction} 

The (1+1)-dimensional nonlinear $\sigma$-model (NLSM) has played important roles in both high energy and condensed matter physics. 
The NLSM shares various common features with the $(3+1)$-dimensional non-Abelian gauge theories, such as the asymptotic freedom~\cite{polyakov-hidden-1977}, dynamical generation of mass gap~\cite{polyakov-interaction-1975}, solitons~\cite{skyrme-nonlinear-1961,belavin-metastable-1975}, and nontrival $\theta$ vacua. 
The NLSM can be further generalized to the $1/N$ expandable $CP^{N-1}$ model~\cite{eichenherr-sun-1978,dadda-1n-1978} which is believed to be relevant to the study of the strong $CP$ problem~\cite{schierholz-towards-1994}.    
Hence, a thorough understanding of the nature of the NLSM can undoubtedly give much insight into the study of the non-Abelian gauge theories in $3+1$ dimensions.  

From the condensed matter side, the (1+1)-dimensional NLSM arises in the context of the Haldane's conjecture~\cite{haldane-nonlinear-1983,haldane-continuum-1983}:
the (1+1)-dimensional $O(3)$ NLSM with $\theta=0$ ($\theta=\pi$) is the low-energy, long-wavelength effective theory for the quantum antiferromagnetic Heisenberg chain with integer (half-integer) spin $S$.
Since the NLSM with $\theta=0$ is known to have exponentially decaying correlations~\cite{polyakov-interaction-1975}, the integer-spin antiferromagnetic (AF) Heisenberg chain is conjectured to be gapped.
Meanwhile, based on known results for the spin-1/2 AF Heisenberg chain, Haldane conjectured~\cite{haldane-continuum-1983} that the $O(3)$ NLSM with $\theta=\pi$ topological term is a massless theory.
Later, by mapping the NLSM with $\theta=\pi$ term to a modified quantum rotor model, Shankar and Read~\cite{shankar-theta-1990} claimed that this model should be massless for all values of the coupling constant.
In the strong-coupling limit, as pointed out by Affleck and Haldane~\cite{affleck-critical-1987,affleck-quantum-1989}, the $O(3)$ NLSM with $\theta=\pi$ is equivalent to the SU(2)$_1$ Wess-Zumino-Novikov-Witten (WZNW) conformal field theory (CFT)~\cite{wess-consequences-1971,novikov-multivalued-1981,witten-nonabelian-1984}, whose central charge is $c=1$. 
In the weak-coupling limit, this model corresponds to two massless bosons, thus has a central charge $c=2$.
In the renormalization group framework, the NLSM with $\theta=\pi$ flows from the unstable fixed point at the weak-coupling limit to the stable fixed point at the strong-coupling limit, and, according to $c$-theorem, the central charge varies monotonically between these two limiting cases. 

From the numerical side, in the study of lattice field theories, the Monte Carlo algorithm has been a standard approach since the beginning of this field.
However, in many cases, the Monte Carlo approach is hindered by the sign problem---more specifically, for example, in the case of NLSM, the straightforward Monte Carlo simulation encounters the sign problem when the $\theta$-term is nonzero. 
Although several approaches, such as the Meron-cluster Monte Carlo algorithm~\cite{bietenholz-meroncluster-1995,bogli-nontrivial-2012,de-forcrand-walking-2012} and the analytic continuation approach based on imaginary $\theta$ simulation data~\cite{azcoiti-critical-2007,alles-mass-2008,azcoiti-critical-2012,alles-behavior-2014}, have been successfully developed to overcome the sign problem for this specific case, these methods are rather specific and cannot be easily extended and applied to other systems. 
On the other hand, in recent years, the tensor network algorithms have achieved rapid development, and have been increasingly applied to the numerical simulation of lattice field theories~\cite{banuls-review-2020,meurice-tensor-2020}. 
Unlike the Monte Carlo approach, the tensor network methods are free from the sign problem, and thus can hopefully be applied to many problems where the Monte Carlo simulation are hindered or even prohibited.
It is then meaningful to develop and test tensor network algorithms for the NLSM with $\theta$-terms.

In this work, we perform a tensor network simulation of the NLSM with $\theta=\pi$ topological term.
Inspired by Ref.~\cite{shankar-theta-1990}, we work in the Hamiltonian formulation and map the NLSM with $\theta=\pi$ term to a modified quantum rotor model where the quantum rotors are decorated with magnetic monopoles.
By representing the modified quantum rotor model in the basis of magnetic monopoles, we obtain its matrix representation which automatically enables MPS-based simulations of this model.
Making use of the recently proposed continuous matrix product operator (cMPO) method~\cite{tang-continuous-2020}, we simulate the finite-temperature properties of the modified quantum rotor model and present clear numerical evidence for its massless nature.
Moreover, we also obtain the central charge as a function of the coupling constant, and compare the result with the field theoretical predictions.  

This paper is organized as follows. 
In Sec.~\ref{sec:hamiltonian-formulation}, we introduce the Hamiltonian formalism, i.e., the quantum rotor model for the NLSM, respectively for $\theta=0$ and $\theta=\pi$ cases. We also introduce the matrix representation for the quantum rotor models. 
In Sec.~\ref{sec:numerical-approach}, we introduce the cMPO approach and its application in the quantum rotor model.
In Sec.~\ref{sec:results}, we show the numerical results which prove the massless nature of the NLSM with $\theta=\pi$.
Finally, Sec.~\ref{sec:summary} summarizes the results and provides some outlook.
In Appendix \ref{app:matrix-representation}, we discuss the matrix representation of the modified quantum rotor model introduced by Sec.~\ref{sec:hamiltonian-formulation} in detail. 
In Appendix \ref{app:proof-mapping}, we present a proof for the mapping between NLSM and quantum rotor models for both $\theta=0$ and $\theta=\pi$ cases. 
Appendix \ref{app:details} includes some details in the numerical simulation.

\section{Hamiltonian formulation} \label{sec:hamiltonian-formulation}

In (1+1) dimensions, the Euclidean action of the $O(3)$ NLSM is given by 
\begin{equation}
  S_0 = \frac{1}{2g^2} \int \mathd x \mathd \tau \, (\partial \boldn)^2 ,
  \label{eq:S0-O3}
\end{equation}
where $\boldn$ is a unit vector that rotates in the three-dimensional space, $g$ is the dimensionless bare coupling constant, and $x$, $\tau$ represent the Euclidean space coordinates.
Due to the possible existence of instantons in this model, one can extend the action by adding a $\theta$-term
\begin{equation}
  S = S_0 + \mathi \frac{\theta}{4 \pi}\int \mathd x \mathd \tau \, \boldn \cdot (\partial_x \boldn \times \partial_\tau \boldn), 
  \label{eq:S-O3-theta}
\end{equation}
where $\theta$ is periodic in $2\pi$. 
The partition function of this model is written in the path integral formulation as 
$Z = \int \mathD \boldn \, \exp (-S [\boldn])$, where, throughout this work, the functional integration $\mathD \boldn$ is defined with respect to the real unit vector field. 
In this section, we will briefly review the Hamiltonian formulation of the NLSM with $\theta=0$ and $\theta=\pi$.

\subsection{Hamiltonian formulation for NLSM with \boldmath$\theta=0$} 
For the NLSM with $\theta=0$, it has been well established that the Hamiltonian formulation is given by the one-dimensional quantum $O(3)$ rotor model on the lattice~\cite{hamer-phases-1978,hamer-strong-1979,shankar-theta-1990,milsted-matrix-2016,bruckmann-O3-2019}
\begin{equation}
  a \hat{H} = \sum_j \frac{\hat{\boldL}_j^2}{2 K} - K \sum_{\< i, j \>} \hat{\boldn}_i \cdot \hat{\boldn}_j ,
  \label{eq:qrm-theta0}
\end{equation}
where $\hat{\boldL}_j$ and $\hat{\boldn}_j$ respectively represent the angular momentum operator and rotor operator on site $j$, $a$ is the lattice spacing, and $K>0$ is a constant. The operators satisfy the following commutation relations:
\begin{equation}
  [\hat{L}^{\mu}_j, \hat{L}^{\nu}_l] = \mathi \varepsilon^{\mu \nu \lambda} \hat{L}^\lambda_j \delta_{jl}, \; 
  [\hat{L}^{\mu}_j, \hat{n}^{\nu}_l] = \mathi \varepsilon^{\mu \nu \lambda} \hat{n}^\lambda_j \delta_{jl}, \;
  [\hat{n}^{\mu}_j, \hat{n}^{\nu}_l] = 0, 
\end{equation}
where $\mu, \nu, \lambda = x, y, z$.
In the low-energy, long-wavelength limit, the field theoretical description of this rotor model is just the $O(3)$ NLSM with $\theta=0$ and $1 / g^2 = K$.

The eigenstate of the rotor operator $\hat{\boldn}$ is parametrized by continuous angle variables, which is not convenient for tensor network simulations.
Instead, a discrete basis is preferred, for which the eigenbasis of the angular momentum operators serves as a natural choice: $\hat{\boldL}^2 |l, m\> = l(l+1) |l, m\>$ and $\hat{L}^z |l, m\> = m |l, m\>$. The quantum numbers $l$ and $m$ take integer values, $l= 0, 1, 2 \ldots$ and $m = -l, -l+1, \ldots, l$.
In this basis, the kinetic term in the Hamiltonian \eqref{eq:qrm-theta0} becomes diagonal. The rotor couplings can be rewritten as $\hat{\boldn}_i \cdot \hat{\boldn}_j = \sum_{\nu \in \{0, \pm\}} \hat{n}^\nu_i \hat{n}^{-\nu}_j$ with $\hat{n}^{\pm} = (\hat{n}^x \pm \mathi \hat{n}^y) / \sqrt{2}$ and $\hat{n}^0 = \hat{n}^z$, whose matrix representation in the angular momentum basis can be obtained by taking $\hat{n}^\nu$ as spherical tensor operators~\cite{bruckmann-O3-2019}.
With the matrix representation of the Hamiltonian, it is then straightforward to represent the partition function as a tensor network or perform ground/excited state calculations via MPS-based methods. 

\subsection{Hamiltonian formulation for NLSM with \boldmath$\theta=\pi$}
In the presence of a nonvanishing $\theta$-term, it is a nontrivial task to incorporate it in the quantum rotor model formulation.
As pointed out in Ref.~\cite{shankar-theta-1990}, by adding a magnetic monopole with magnetic charge $q=1/2$ at the center of the rotor and setting nearest-neighbor couplings to be antiferromagnetic, the low-energy effective theory becomes a NLSM with $\theta=\pi$.
In the Hamiltonian formulation, the presence of the magnetic monopole modifies the definition of the angular momentum operator $\hat{\boldL}$, and the Hamiltonian becomes~\cite{shankar-theta-1990}
\begin{equation}
  a \hat{H} = \sum_j \frac{(\hat{\boldL}'_j)^2}{2 K} + K \sum_{\< i, j \>} \hat{\boldn}_i \cdot \hat{\boldn}_j \,.
  \label{eq:qrm-thetapi}
\end{equation}
In spatial coordinates, the modified angular momentum operator is defined by $\hat{\boldL}' = \boldn \times (-\mathi \nabla - \tmmathbf{A}) - \boldn$, where $\boldn$ is the unit vector pointing at the direction of the rotor, and $\tmmathbf{A}$ is the vector potential describing the magnetic field generated by the magnetic monopole (see, e.g., Ref.~\cite{wu-dirac-1976} for more details).
The low-energy physics of this model is described by the NLSM with $\theta=\pi$, whose coupling constant satisfies $1/g^2 = K$.

To find the matrix representation of the Hamiltonian~\eqref{eq:qrm-thetapi}, we make use of the eigenbasis of the modified angular momentum operator --- the monopole harmonics~\cite{wu-dirac-1976,wu-some-1977}.
It is known that $\hat{\boldL}'$ still satisfies the angular momentum commutation relations~\cite{wu-dirac-1976}, and the eigenbasis of monopole harmonics can be labeled by well-defined angular momentum quantum numbers $(l, m)$, which satisfies $(\hat{\boldL}')^2 |q,l,m\> = l(l+1) |q,l,m\>$ and $(\hat{L}')^z = m|q,l,m\>$. Here, the quantum number $q$ denotes the magnetic charge at the center of the rotor and hence takes the value $q=1/2$, which distinguishes itself from the ordinary spherical harmonics with $q=0$.
In the presence of the magnetic charge $q=1/2$, $l$ and $m$ can only take half-integer values, $l=1/2, 3/2, 5/2, \ldots$ and $m=-l, -l+1, \ldots, l$. 
Making use of this angular momentum eigenbasis, the matrix representation of the Hamiltonian in \Eq{eq:qrm-thetapi} is similar to the case with $\theta=0$ --- 
the kinetic term is diagonal, and the matrix representation for the rotor coupling term can be evaluated by using the properties of spherical tensor operators, the details of which are included in Appendix \ref{app:matrix-representation}. 
%We include the details for the matrix representation in Appendix \ref{app:matrix-representation}.
Based on the properties of the monopole harmonics, we also provide a proof for the mapping between the lattice Hamiltonian and the continuous theory of the NLSM with $\theta=\pi$ in Appendix \ref{app:proof-mapping}.

In practical simulations, we need to truncate the physical Hilbert space at each site.
The most natural choice is to choose a maximally allowed angular momentum quantum number $l_{\mathrm{max}}$ and drop the states with $l > l_{\mathrm{max}}$.
For the Hamiltonian in Eq.~\eqref{eq:qrm-thetapi}, one can infer that this truncation scheme is effective only when the constant $K$ is small, i.e., near the strong coupling limit. 
An interesting limit is $K \rightarrow 0$, where one can choose $l_{\mathrm{max}} = 1/2$, and the modified quantum rotor model reduces to the $S=1/2$ antiferromagnetic Heisenberg chain. 
For large values of $K$, in principle, one has to use large enough $l_{\mathrm{max}}$ to obtain quantitatively accurate results.  

\section{Tensor network approach to the modified quantum rotor model} \label{sec:numerical-approach}

From the numerical side, we make use of the recently developed cMPO method~\cite{tang-continuous-2020} to study the finite temperature properties of the (modified) quantum rotor model in Eq.~\eqref{eq:qrm-thetapi}.
The reason for using this approach is twofold.
First, since the theory is expected to be massless for all choices of the coupling constant, working at the finite temperature can help reduce the requirement on the bond dimensions compared to ground-state simulations in the thermodynamic limit~\cite{znidaric-complexity-2008,barthel-one-dimensional-2017}. 
It also allows us to adjust the temperature for different choices of the coupling constant. 
Second, the cMPO approach works in the continuous time limit, which automatically eliminates the discretization error in the imaginary time direction.

In this section, we will briefly review the cMPO approach, and introduce the cMPO formulation for the modified quantum rotor model defined in Eq.~\eqref{eq:qrm-thetapi}. 
We will also discuss two key properties of the cMPO for this model: (i) Hermiticity, which enables a direct global optimization during the simulation; (ii) Symmetry, which is inherited from the Hamiltonian and allows us to reduce the computational cost in numerical simulations. 

\subsection{Brief review of the cMPO approach}

\begin{figure}[t]
  \centering
  \resizebox{\columnwidth}{!}{\includegraphics{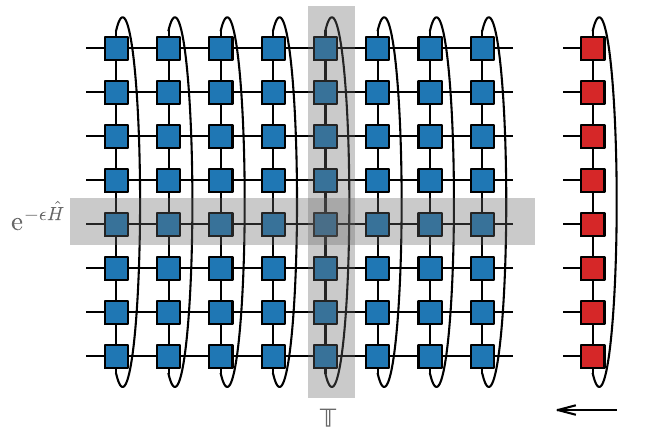}}
  \caption{The tensor network representation for the partition function $Z = \tr \, \mathrm{e}^{-\beta \hat{H}}$ and the (right) boundary MPS. 
  The blue blocks represent the local tensor $T$, whose vertical and horizontal legs are respectively referred to as physical and virtual bonds.
  The red blocks represent the local tensor $T_{\psi}$ of the boundary cMPS.
  } \label{fig:concept}
\end{figure}

%First, we briefly review the cMPO method.
The cMPO approach is based on the observation that there exists a compact MPO representation for the infinitesimal time evolution operator $\exp(-\epsilon \hat{H})$ when we only consider up to the first order in $\epsilon$~\cite{zaletel-time-2015}.
The neglected higher order terms of $\epsilon$ will not incur any errors since we will take the $\epsilon \rightarrow 0$ limit.
From this MPO one can build the tensor network representation for the partition function $Z = \tr \, \mathrm{e}^{-\beta \hat{H}}$ (see Fig.~\ref{fig:concept}). 
The local tensor $T$ in the MPO can be expressed as 
\begin{equation}
  T =
\left( \begin{array}{cc}
  I + \epsilon \sfQ & \sqrt{\epsilon} \sfR \\
  \sqrt{\epsilon} \sfL & \sfP 
  \end{array} \right),
  \label{eq:cmpo-general}
\end{equation}
where $\sfQ$ is an operator-valued scalar, $I$ is the identity operator, $\sfL$ and $\sfR$ are operator-valued vectors (not to be confused with the angular momentum operator $\hat{\boldL}$), and $\sfP$ is a matrix consisting of operators.
The operators contained in $\sfQ$, $\sfL$, $\sfR$, and $\sfP$ are operators acting on the physical Hilbert space, which all come from the Hamiltonian:
$\sfQ$ corresponds to local terms, $\sfL$ and $\sfR$ encode nearest-neighbor interactions, and $\sfP$ comes from longer-range interactions.
The physical dimension is thus the dimension of the physical Hilbert space at each site. 
The virtual bond dimension $D=d+1$, where $d$ is the dimension of vectors $\sfL$ and $\sfR$.

%Next, note that the partition function can be expressed as $Z = \tr [\exp(-\epsilon \hat{H})^{\beta / \epsilon}]$.
%As $\exp(-\epsilon \hat{H})$ can be represented as an MPO, this expression leads to a tensor network representation of the partition function $Z$ by stacking $\beta/\epsilon$ layers of MPOs together \hhtc{Add a figure for the tensor network?}. %[see Fig.~\ref{fig:concept}(a)].
Next, as shown in Fig.~\ref{fig:concept}, the tensor network for the partition function $Z$ can be formed by stacking $\beta/\epsilon$ layers of the MPOs together.  
In the thermodynamic limit, i.e., $L \rightarrow \infty$, this tensor network can be efficiently contracted using the idea of transfer matrix~\cite{bursill-density-1996,wang-transfer-1997,xiang-thermodynamics-1998}. 
The transfer matrix $\mathbb{T}$ refers to the column of tensors in the tensor network, which is also an MPO, and we can approximate its dominant eigenvector with an MPS, which is referred to as the boundary MPS (see Fig.~\ref{fig:concept}).
More specifically, here, as we take the continuous time limit $\epsilon \rightarrow 0$, the MPO representation for the transfer matrix becomes continuous (hence the name cMPO), and the corresponding boundary MPS becomes a continuous MPS (cMPS).
Since the cMPO is uniform with the periodic boundary condition (along the imaginary time direction), it is natural to use a uniform cMPS $|\psi\>$, parametrized by the local tensor
\begin{equation}
  T_{\psi} = 
\left( \begin{array}{c}
  I_{\psi} + \epsilon \sfQ_\psi  \\
  \sqrt{\epsilon} \sfR_\psi 
  \end{array} \right)
  \label{eq:cmps}
\end{equation}
as the boundary cMPS.
Similar to Eq.~\eqref{eq:cmpo-general}, $\sfQ_\psi$ and $\sfR_{\psi}$ in Eq.~\eqref{eq:cmps} correspond to an operator and a vector of operators, respectively. 
The operators contained in $T_{\psi}$ are parametrized by matrices, whose dimension is the bond dimension of the cMPS. 
From the boundary cMPS, one can further extract the thermodynamic properties of the system.

To obtain the boundary cMPS $|\psi\>$, one can directly minimize the free energy density  
\begin{equation}
 f = -\frac{1}{\beta} \ln (\<\psi|\mathbb{T}|\psi\> / \<\psi|\psi\>)
 \label{eq:free-energy}
\end{equation}
if the cMPO $\mathbb{T}$ is Hermitian. 
In general, $\mathbb{T}$ is non-Hermitian, and $|\psi\>$ has to be optimized by the power method, i.e., by repeatedly acting $\mathbb{T}$ on a trial solution for $|\psi\>$ and compressing its bond dimension. At each iteration step, the compression of $\mathbb{T} |\psi\>$ into the cMPS $|\phi\>$ (with a smaller bond dimension) is again a variational optimization process that maximizes the fidelity $\mathcal{F} = \< \phi | \mathbb{T} |\psi \>/\sqrt{\<\phi | \phi\> }$, where we have dropped a constant factor $1/\sqrt{\< \psi|\mathbb{T}^\dagger \mathbb{T}| \psi \>}$ for simplicity.

\subsection{cMPO formulation for the modified quantum rotor model} \label{sec:cmpo-formulation-qrm}

For the modified quantum rotor Hamiltonian in \Eq{eq:qrm-thetapi}, the local tensor $T$ is given by 
\begin{equation}
\left( \begin{array}{c|ccc}
   I + \epsilon (\hat{\boldL}')^2/2K & \sqrt{\epsilon K} \hat{n}^+ & \sqrt{\epsilon K} \hat{n}^- & \sqrt{\epsilon K} \hat{n}^0 \\
    \hline
   -\sqrt{\epsilon K} \hat{n}^- &  &  & \\
   -\sqrt{\epsilon K} \hat{n}^+ &  &  & \\
   -\sqrt{\epsilon K} \hat{n}^0 &  &  & 
  \end{array} \right).
  \label{eq:cmpo-o3-nlsm}
\end{equation}
One can easily identify the contents of $\sfQ$, $\sfL$, $\sfR$, and $\sfP$ in Eq.~\eqref{eq:cmpo-general}.  
It is worth mentioning that this cMPO is Hermitian.
To see this, we can perform a unitary transformation described by $\hat{U}=\exp(\mathi \pi (\hat{L}')^y]$ to all the operators contained in the local tensor $T$.
Using the commutation relations between $\hat{\boldL}'$ and $\hat{\boldn}$, we find
\begin{equation}
  \hat{U} (\hat{\boldL}')^2 \hat{U}^\dagger = (\hat{\boldL}')^2, \;
  \hat{U} \hat{n}^0 \hat{U}^\dagger = - \hat{n}^0, \;
  \hat{U} \hat{n}^\pm \hat{U}^\dagger = - \hat{n}^\mp.
  \label{eq:rotation-of-n}
\end{equation}
On the one hand, according to Eq.~\eqref{eq:rotation-of-n}, $\hat{U}$ switches the contents of $\sfL$ and $\sfR$, and since $\sfP$ is empty, it effectively switches the left and right bonds of the local tensor $T$ [see Eq.~\eqref{eq:cmpo-general} and Fig.~\ref{fig:concept}].
When viewed as a large matrix, the cMPO $\mathbb{T}$ becomes its own transpose (and also its own Hermitian conjugate, as $\mathbb{T}$ is a real matrix) after this unitary transformation.
On the other hand, since the physical bonds of the local tensors in cMPO are all contracted (see Fig.~\ref{fig:concept}), this unitary transformation is merely a gauge transformation and leaves the cMPO unchanged.
Therefore, the cMPO $\mathbb{T}$ is Hermitian, which allows us to directly optimize the boundary cMPS by variationally minimizing the free energy in \Eq{eq:free-energy}.

In our calculations, before variationally minimizing the free energy, we perform a few power method steps to obtain a good initialization for the variational optimization.
To help stabilize the power method procedure, we apply the unitary transformation $\exp[\mathi \pi (\hat{L}')^z]$ on every second site, such that the rotor couplings in $x$ and $y$ directions in the Hamiltonian~\eqref{eq:qrm-thetapi} become ferromagnetic~\footnote{The relation between the stability of the power method and the antiferromagnetic couplings in the Hamiltonian is beyond the scope of this paper and requires further investigations. Here, the rotation process is merely a numerical trick and does not affect the physical results.},
\begin{equation}
  \hat{H} = \sum_j \frac{({\hat{\boldL}}'_j)^2}{2 K} + K \sum_{\< i, j \>} ( - \hat{n}_i^{+} \hat{n}_j^{-} - \hat{n}_i^{-} \hat{n}_j^{+} + \hat{n}_i^0 \hat{n}_j^0).
  \label{eq:rotated-rotor-model}
\end{equation}
%This gauge transformation can be achieved by alternatively rotating the rotors by an angle $\pi$ around the $z$ axis, i.e., by the unitary transformation $\exp(\mathi \pi \hat{L}^z)$.
For this ``rotated'' Hamiltonian, the local tensor now reads
\begin{equation}
\left( \begin{array}{c|ccc}
   I + \epsilon (\hat{\boldL}')^2/2K & \sqrt{\epsilon K} \hat{n}^+ & \sqrt{\epsilon K} \hat{n}^- & \sqrt{\epsilon K} \hat{n}^0 \\
    \hline
   \sqrt{\epsilon K} \hat{n}^- &  &  & \\
   \sqrt{\epsilon K} \hat{n}^+ &  &  & \\
   -\sqrt{\epsilon K} \hat{n}^0 &  &  & 
  \end{array} \right) .
  \label{eq:cmpo-o3-nlsm-rotated}
\end{equation}
The Hermiticity of the cMPO can be proven in an analogous way. 

\subsection{\boldmath$U(1)$ symmetry of the cMPO}

Although the Hamiltonian in \Eq{eq:rotated-rotor-model} is SO(3) symmetric, we shall just use its $U(1)$ subgroup (i.e., rotational invariance around the $z$ axis) in our tensor network simulations.
Mathematically, the $U(1)$ symmetry is generated by $\hat{U}(\theta) = \exp[\mathi \theta (\hat{L}')^z]$, with the operators transforming as 
\begin{equation}
  \hat{U}(\theta) (\hat{\boldL}')^2 \hat{U}(\theta)^\dagger = (\hat{\boldL}')^2, \quad
  \hat{U}(\theta) \hat{n}^\nu \hat{U}(\theta)^\dagger = \mathe^{\mathi \nu \theta} \hat{n}^\nu,
  \label{eq:U1-rotation}
\end{equation}
where $\nu = 0, \pm$. 
Equation \eqref{eq:U1-rotation} not only proves the $U(1)$ invariance of the Hamiltonian, but also indicates that the $U(1)$ symmetry can be encoded into the cMPO. 
Combining Eq.~\eqref{eq:cmpo-o3-nlsm-rotated} and Eq.~\eqref{eq:U1-rotation}, we get 
\begin{equation}
  \begin{matrix}
  \includegraphics{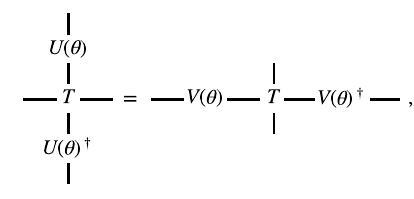}
  \end{matrix}
  \label{eq:U1-rotation-cmpo-tensor}
\end{equation}
where, the same as Fig.~\ref{fig:concept}, the vertical and horizontal lines respectively correspond to the physical and virtual indices of the local tensor $T$. $U(\theta)$ is the matrix representation of $\hat{U}(\theta)$. 
On the virtual bonds, $V(\theta) = \exp(\mathi \theta Z)$, where $Z = \mathrm{diag}(0, 1, -1, 0)$.
As shown in Fig.~\ref{fig:concept}, the physical indices of local tensors in the cMPO are all contracted, and the left-hand side of Eq.~\eqref{eq:U1-rotation-cmpo-tensor} is thus a gauge transformation that leaves the cMPO invariant. 
Meanwhile, the right-hand side of Eq.~\eqref{eq:U1-rotation-cmpo-tensor} gives rise to a global $U(1)$ rotation, which is described by $V(\theta)$ on each virtual bond.
Therefore, we can take $V(\theta)$ as a symmetry transformation of the cMPO. 
Based on this fact, we can correspondingly construct a $U(1)$-symmetric boundary cMPS, which contains block structures and helps lower down the computational cost.
The construction of the $U(1)$-invariant boundary cMPS follows the general rules to construct symmetric MPS~\cite{peres-garcia-string-2008,sanz-matrix-2009,singh-tensor-2010,singh-tensor-2011}. More specifically, we build the boundary cMPS from the local tensors satisfying
\begin{equation}
  \begin{matrix}
  \includegraphics{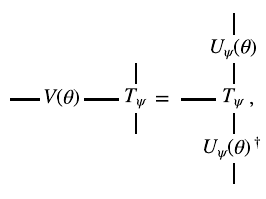}
  \end{matrix}
  \label{eq:U1-symmetric-cmps-tensor}
\end{equation}
up to a phase factor.
Here, $T_{\psi}$ represents the cMPS local tensor, and $U_{\psi}(\theta)$ denotes the $U(1)$ rotation acting on the internal indices of the cMPS.
The degeneracy sectors on the cMPS vertical bond are determined dynamically, the details of which are included in Appendix \ref{app:u1}.
%We include more details about the $U(1)$-symmetric boundary cMPS in the Appendix \ref{app:u1}.

\section{Results} \label{sec:results}

In this section, we describe our numerical results, which provide numerical evidence of the massless nature of the NLSM with $\theta=\pi$. 
More specifically, we demonstrate the results for the free energy density and the bipartite entanglement entropy of the boundary cMPS and compare them with the predictions of CFT.  

In our numerical simulation, we calculate the finite temperature properties of the modified quantum rotor model.
We perform simulations from $K=1.0$ to $K=6.0$ in order to cover a fairly large range of values for the coupling constant.
The range of temperatures varies with $K$. For each choice of $K$, the temperature $T$ ranges from $K/300$ to $K/60$, since the energy scale increases with $K$.
The maximal angular momentums are chosen to be $l_{\mathrm{max}} = 1/2, 3/2, 5/2$.   
The bond dimension of the boundary cMPS ranges among $\chi=12, 18, 24, 30$, and we extrapolate the results to an infinite bond dimension (see Appendix \ref{app:extrapolation} for details).

Our code implementation is publicly available at \footnote{See \url{https://github.com/tensorBFS/U1cMPO} for code implementation in Julia}.

\subsection{Universal correction to free energy}

For one-dimensional quantum systems described by CFT, a universal finite-size correction to the free energy appears at low temperature~\cite{affleck-universal-1986,blote-conformal-1986}
\begin{equation}
  F / L = f_0 - \frac{\pi c T^2}{6 v},
  \label{eq:free-energy-corr}
\end{equation}
where $f_0$ is the free energy density at zero temperature, $c$ is the central charge, and $v$ is the effective ``velocity of light'' in the theory. Equation \eqref{eq:free-energy-corr} predicts that the specific heat is linear in $T$. This property is in sharp contrast to that of gapped systems, where the free energy manifests an exponential scaling at low temperatures.

\begin{figure}[!htb]
  \centering
  \resizebox{\columnwidth}{!}{\includegraphics{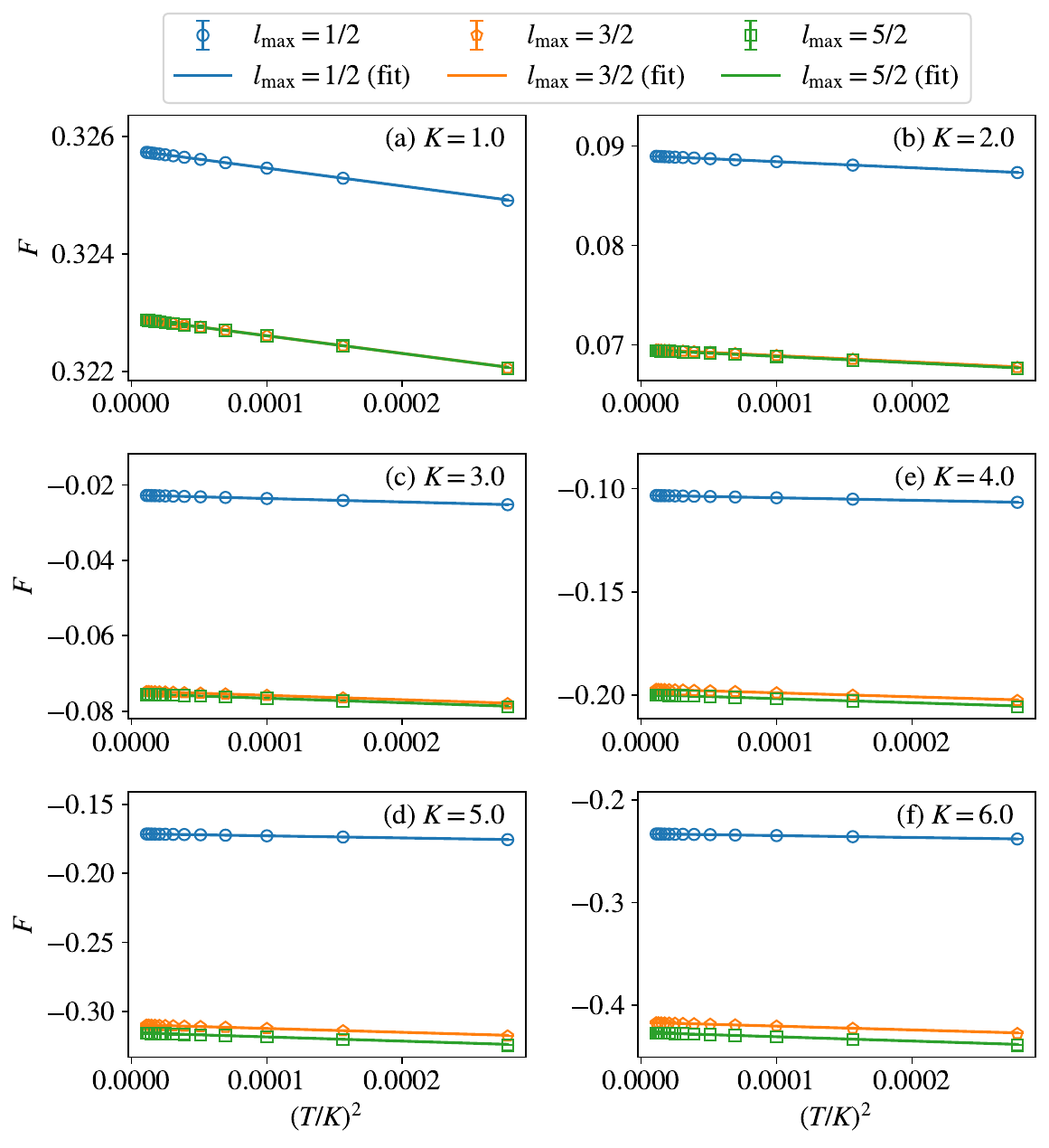}}
  \caption{The free energy density with respect to $(T/K)^2$ for the modified quantum rotor model from $K=1.0$ to $K=6.0$. 
  Results obtained with different $l_{\mathrm{max}}$ are marked with different colors.
  The error bars come from the uncertainty in the extrapolation to the infinite bond dimension. 
  The linear fitting is performed for $100 \leq K/T \leq 300$.}
  \label{fig:Ffits}
\end{figure}

From our calculation results, we verify the massless nature of the system by comparing with \Eq{eq:free-energy-corr}.
Figure \ref{fig:Ffits} shows the free energy density as a function of $(T/K)^2$ for different choices of $K$ and the linear fitting of the data to Eq.~\eqref{eq:free-energy-corr}. 
The linear fitting is performed within the range $K/300 \leq T \leq K/100$, the detailed results of which are shown in Table \ref{tab:free-energy-fits}.
For all values of $K$ that we consider, we observe a clear linear dependence of the free energies on $T^2$, which perfectly coincides with the prediction of Eq.~\eqref{eq:free-energy-corr}.
For small values of $K$, the results quickly converge with respect to $\lmax$.  
As $K$ increases, the results for different $\lmax$'s gradually deviate from each other, since the angular momentum basis becomes less effective as the system approaches the weak coupling limit.
Nonetheless, they show a tendency of convergence and still serve as qualitative evidence for the massless nature of this model. 

\begin{table}[]
\centering
\begin{threeparttable}
\begin{tabular}{p{0.04\textwidth}>{\centering}p{0.19\textwidth}>{\centering\arraybackslash}p{0.22\textwidth}}%{@{}ccc@{}}
\toprule
\multicolumn{3}{c}{$K=1.0$}                        \\
  $l_{\mathrm{max}}$   & $-\pi c K^2 / 6v$ & $f_0$ \\
$1/2$   & -3.03879 $\pm$ 0.00063   & 0.32576 $\pm$ 5.86395 $\times 10^{-8}$   \\
$3/2$   & -3.00617 $\pm$ 0.19215   & 0.32291 $\pm$ 1.58070 $\times 10^{-5}$   \\
$5/2$   & -3.00531 $\pm$ 0.20338   & 0.32291 $\pm$ 1.68733 $\times 10^{-5}$   \\[4pt] \hline
\multicolumn{3}{c}{$K=2.0$\Tstrut}                        \\
  $l_{\mathrm{max}}$   & $-\pi c K^2 / 6v$ & $f_0$ \\
$1/2$   & -1.51970 $\pm$ 0.00032   & 0.08903 $\pm$ 1.16654 $\times 10^{-7}$   \\
$3/2$   & -1.59886 $\pm$ 0.08748   & 0.06955 $\pm$ 3.23684 $\times 10^{-5}$   \\
$5/2$   & -1.62788 $\pm$ 0.10347   & 0.06948 $\pm$ 3.03028 $\times 10^{-5}$   \\[4pt] \hline
\multicolumn{3}{c}{$K=3.0$\Tstrut}                        \\
  $l_{\mathrm{max}}$   & $-\pi c K^2 / 6v$ & $f_0$ \\
$1/2$   & -1.01339 $\pm$ 0.00019   & -0.02271 $\pm$ 1.59098 $\times 10^{-7}$   \\
$3/2$   & -1.28895 $\pm$ 0.05683   & -0.07466 $\pm$ 4.34769 $\times 10^{-5}$   \\
$5/2$   & -1.31000 $\pm$ 0.09805   & -0.07541 $\pm$ 6.34446 $\times 10^{-5}$   \\[4pt] \hline
\multicolumn{3}{c}{$K=4.0$\Tstrut}                        \\
  $l_{\mathrm{max}}$   & $-\pi c K^2 / 6v$ & $f_0$ \\
$1/2$   & -0.75967 $\pm$ 0.00014   & -0.10320 $\pm$ 2.10380 $\times 10^{-7}$   \\
$3/2$   & -1.19251 $\pm$ 0.05365   & -0.19685 $\pm$ 7.11228 $\times 10^{-5}$   \\
$5/2$   & -1.24382 $\pm$ 0.10488   & -0.19956 $\pm$ 1.28900 $\times 10^{-4}$   \\[4pt] \hline
\multicolumn{3}{c}{$K=5.0$\Tstrut}                        \\
  $l_{\mathrm{max}}$   & $-\pi c K^2 / 6v$ & $f_0$ \\
$1/2$   & -0.60774 $\pm$ 0.00011   & -0.17119 $\pm$ 2.49746 $\times 10^{-7}$   \\
$3/2$   & -1.11522 $\pm$ 0.05148   & -0.30964 $\pm$ 1.05917 $\times 10^{-4}$   \\
$5/2$   & -1.21338 $\pm$ 0.11224   & -0.31553 $\pm$ 2.14151 $\times 10^{-4}$   \\[4pt] \hline
\multicolumn{3}{c}{$K=6.0$\Tstrut}                        \\
  $l_{\mathrm{max}}$   & $-\pi c K^2 / 6v$ & $f_0$ \\
$1/2$   & -0.50583 $\pm$ 0.00010   & -0.23293 $\pm$ 3.57072 $\times 10^{-7}$   \\
$3/2$   & -1.02611 $\pm$ 0.04977   & -0.41645 $\pm$ 1.46535 $\times 10^{-4}$   \\
$5/2$   & -1.15988 $\pm$ 0.09968   & -0.42632 $\pm$ 2.86595 $\times 10^{-4}$   \\[4pt] 
\bottomrule 
\end{tabular}
\end{threeparttable}
\caption{Fitting results of the free energy data.
For each $K$ and $l_{\mathrm{max}}$, we list the estimation and the error for the slope $-\pi c K^2 / 6 v$ and the intercept $f_0$. 
The errors originate from both the uncertainty in the extrapolation of the original data to infinite bond dimension and the uncertainty in the linear fitting.
The intercept $f_0$ results also serve as estimations of the ground state energies.
} \label{tab:free-energy-fits}
\end{table}

\subsection{Temporal entanglement entropy}

For systems described by CFT, the temporal direction and the spatial direction are equivalent, and the boundary cMPS can be viewed as the ground state of the ``temporal Hamiltonian'' which is described by the same CFT. 
Therefore, the bipartite entanglement entropy of the boundary cMPS is another important indicator of the massless nature of the system, as it satisfies~\cite{holzhey-geometric-1994,vidal-entanglement-2003,calabrese-entanglement-2004,calabrese-entanglement-2009}
\begin{equation}
  S = \frac{c}{3} \ln \beta + S_0,
  \label{eq:bipartite-entanglement-scaling}
\end{equation} 
where $c$ is the central charge, and $S_0$ is a nonuniversal term.
By fitting the entanglement entropy with respect to the inverse temperature $\beta$, we can extract the central charge of the system. 

\begin{figure}[!htb]
  \centering
  \resizebox{0.5\columnwidth}{!}{\includegraphics{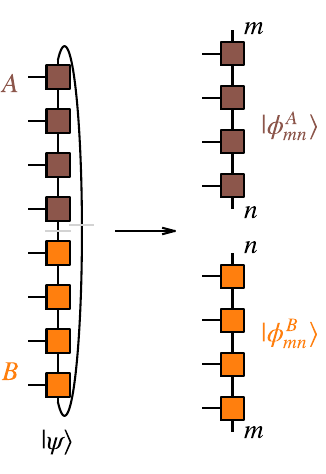}}
  \caption{Dividing the cMPS $|\psi\rangle$ into two parts with equal length $A$ and $B$.}
  \label{fig:bipartite}
\end{figure}

The bipartite entanglement entropy in the uniform temporal cMPS is calculated as follows. 
We divide the cMPS by an equal bipartition, and calculate the von Neumann entropy between the two subsystems $A$ and $B$ (see Fig.~\ref{fig:bipartite}), 
\begin{equation}
  S=-\tr(\rho^A \ln \rho^A), \text{where } \rho^A = \tr_B |\psi\>\<\psi|. 
  \label{eq:vonNeumann}
\end{equation}
From the bipartition, the boundary cMPS can naturally be written as 
\begin{equation}
  |\psi\> = \sum_{m,n} |\phi^A_{mn} \> |\phi^B_{mn}\>, 
\end{equation}
where $m,n$ denote virtual indices through the cuts.
To transform $\{|\phi^{A(B)}_{mn} \>\}$ to an orthonormal basis, we consider the matrix $M^{A(B)}_{m'n',mn} = \< \phi^{A(B)}_{m'n'} | \phi^{A(B)}_{mn} \>$ and its eigen-decomposition 
\begin{equation}
  M^{A(B)} = (U^{A(B)})^\dagger \Lambda^{A(B)} U^{A(B)}.
\end{equation}
One can easily verify that 
\begin{equation}
|\psi^{A(B)}_i\> = \sum_{mn} \frac{(U^{A(B)}_{i,mn})^\ast}{\sqrt{\Lambda^{A(B)}_i}} |\phi^{A(B)}_{mn} \> 
\end{equation}
satisfies $\< \psi^{A(B)}_i | \psi^{A(B)}_{j} \> = \delta_{ij}$.
In this new basis, the cMPS expresses as
\begin{equation}
  |\psi \> = \sum_{i,j} [\sqrt{\Lambda^A} U^A (U^B)^T \sqrt{\Lambda^B}]_{i,j} |\psi_i^A\rangle |\psi_j^B\rangle,
\end{equation}
and the reduced density matrix is given by
\begin{equation}
  \rho^A = \sqrt{\Lambda^A} U^A (M^B)^\ast (U^A)^\dagger\sqrt{\Lambda^A}.
\end{equation} 
Combining this equation with Eq.~\eqref{eq:vonNeumann}, one can evaluate the entanglement entropy.

\begin{figure}[!htb]
  \centering
  \resizebox{\columnwidth}{!}{\includegraphics{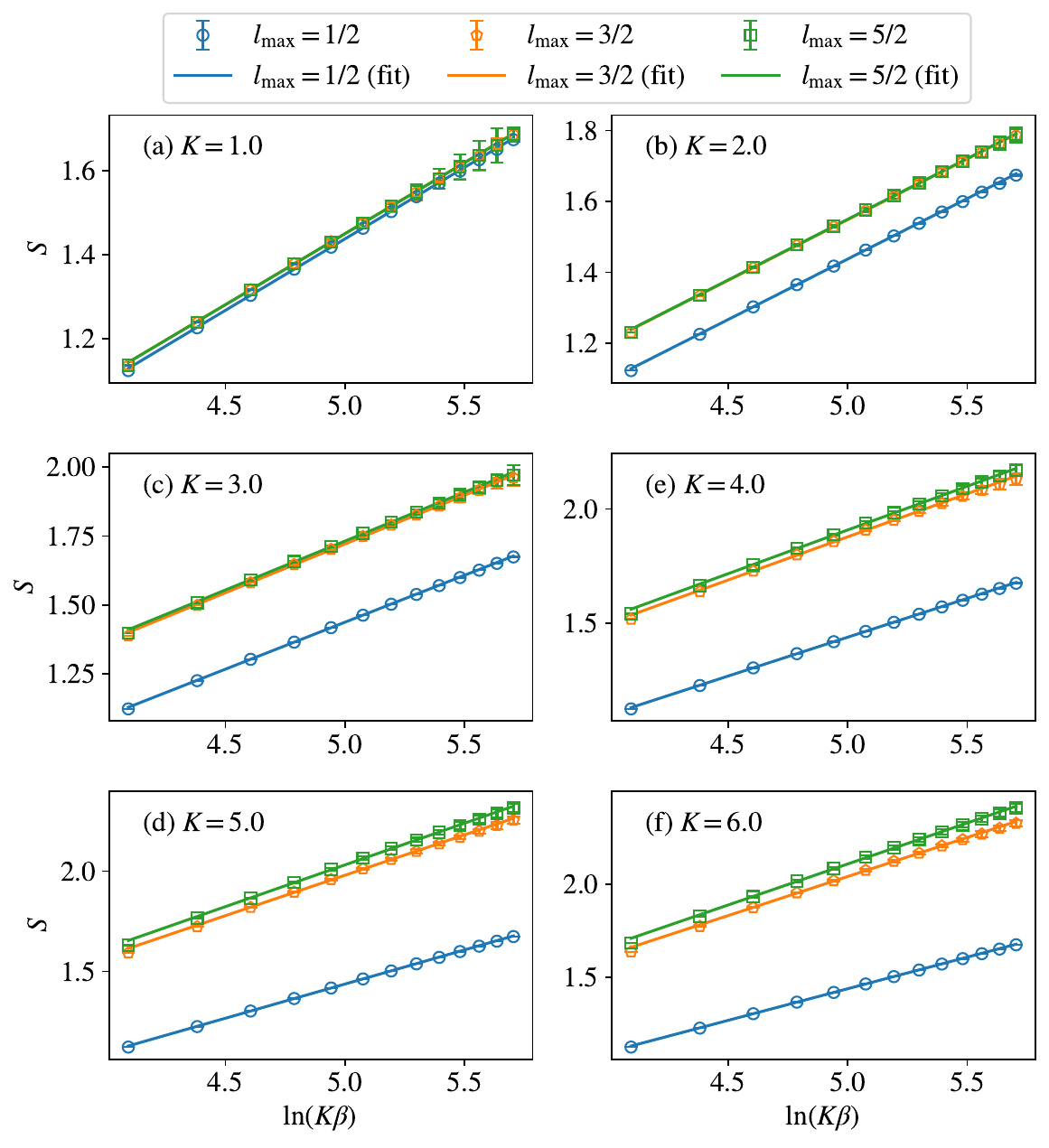}}
  \caption{The bipartite entanglement entropy with respect to $\ln(K \beta)$ for different values of $K$. Results with different choices of $\lmax$ are marked with different colors.
  The error bars come from the uncertainty in the extrapolation to infinite bond dimension. 
  The linear fitting is performed for $100 \leq K\beta \leq 300$.} 
  \label{fig:Sfits}
\end{figure}

In Fig.~\ref{fig:Sfits}, for different values of $K$ and $\lmax$, we show the bipartite entanglement entropy of the boundary cMPS as a function of $\ln(K\beta)$  and the linear fitting of the data to Eq.~\eqref{eq:bipartite-entanglement-scaling}. 
The linear fitting results are shown in Table \ref{tab:entanglement-fits}.
As the parameter $K$ increases, results for different $\lmax$ gradually deviate from each other, which is similar to the results of the free energy.
Nonetheless, these results still show a tendency to converge as $\lmax$ increases, and, at least qualitatively, from these results we can confirm the linear relation between the bipartite entanglement entropy and $\ln \beta$, which verifies Eq.~\eqref{eq:bipartite-entanglement-scaling}. 

\begin{figure}[!htb]
  \centering
  \resizebox{\columnwidth}{!}{\includegraphics{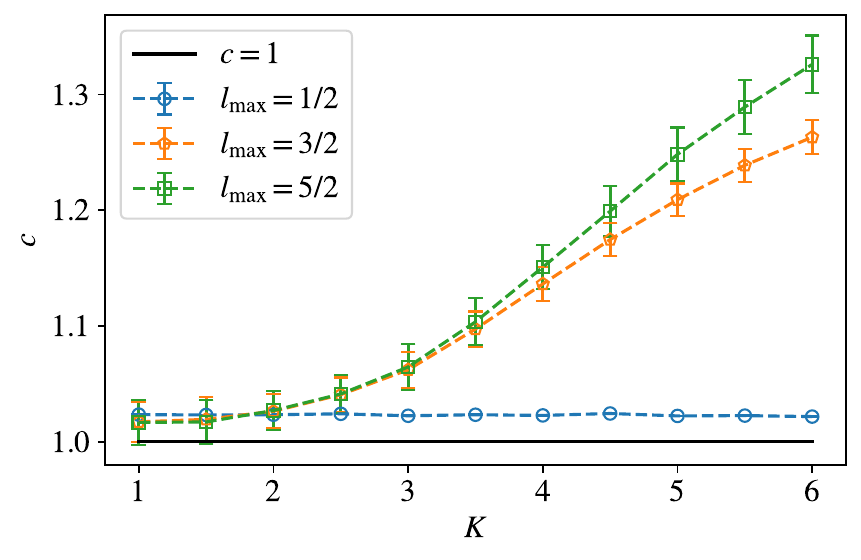}}
  \caption{The dependence of the central charge on the coupling strength $K$ for different choices of $l_{\mathrm{max}}$.
  In the determination of errors in $c$, both the uncertainty in the extrapolation of bipartite entanglement entropy data to infinite bond dimension and the uncertainty in the linear fitting are taken into account. 
  The horizontal black solid line marks $c=1$.} 
  \label{fig:central-charge}
\end{figure}

Furthermore, with Eq.~\eqref{eq:bipartite-entanglement-scaling}, we can obtain the central charge $c$ from the linear regression of the bipartite entanglement data.
As pointed out in Refs.~\cite{affleck-critical-1987,shankar-theta-1990}, in the renormalization group framework, the NLSM with $\theta=\pi$ flows from the unstable fixed point at the weak-coupling limit ($g=0$ and central charge $c=2$) to the stable WZNW fixed point at the strong-coupling limit ($g=\infty$ and $c=1$), and the central charge varies monotonically between the two fixed points according to the $c$-theorem.
Figure \ref{fig:central-charge} shows the numerical results of the central charge.
For the special case of $\lmax=1/2$, the system reduced to the antiferromagnetic Heisenberg model, which is well known to be described by the SU(2)$_1$ WZNW model~\cite{affleck-critical-1987,affleck-critical-1989}.
As shown in Fig.~\ref{fig:central-charge}, for $\lmax=1/2$, and for all values of $K$, the numerical result indeed gives $c \approx 1$, and it can get closer to $c=1$ if one pushes the calculation to lower temperatures. 
For $\lmax = 3/2, 5/2$, the numerical result for central charge shows a monotonic tendency with respect to $K$ and approaches $c\approx 1$ as $K$ becomes small. 
As is similar to the previous results, the result reaches good convergence at $\lmax=5/2$ when $K$ is small ($K\leq 3.0$), but only serves as qualitative evidence for larger values of $K$.  

\begin{table}[]
\centering
\begin{threeparttable}
\begin{tabular}{p{0.04\textwidth}>{\centering}p{0.2\textwidth}>{\centering\arraybackslash}p{0.2\textwidth}}%{@{}ccc@{}}
\toprule
\multicolumn{3}{c}{$K=1.0$}                        \\
  $l_{\mathrm{max}}$   & $c$ & $S_0$ \\
$1/2$   & 1.02355 $\pm$ 0.00018 & -0.26806 $\pm$ 0.00028   \\
$3/2$   & 1.01731 $\pm$ 0.01709 & -0.24520 $\pm$ 0.02682   \\
$5/2$   & 1.01684 $\pm$ 0.01935 & -0.24454 $\pm$ 0.03019   \\[4pt] \hline
\multicolumn{3}{c}{$K=2.0$\Tstrut}                        \\
  $l_{\mathrm{max}}$   & $c$ & $S_0$ \\
$1/2$   & 1.02346 $\pm$ 0.00015 & -0.26791 $\pm$ 0.00024    \\
$3/2$   & 1.02650 $\pm$ 0.01485 & -0.16209 $\pm$ 0.02337    \\
$5/2$   & 1.02711 $\pm$ 0.01699 & -0.16200 $\pm$ 0.02710    \\[4pt] \hline
\multicolumn{3}{c}{$K=3.0$\Tstrut}                        \\
  $l_{\mathrm{max}}$   & $c$ & $S_0$ \\
$1/2$   & 1.02258 $\pm$ 0.00019 & -0.26646 $\pm$ 0.00031    \\
$3/2$   & 1.06227 $\pm$ 0.01535 & -0.04893 $\pm$ 0.02423    \\
$5/2$   & 1.06470 $\pm$ 0.01976 & -0.04240 $\pm$ 0.03167    \\[4pt] \hline
\multicolumn{3}{c}{$K=4.0$\Tstrut}                        \\
  $l_{\mathrm{max}}$   & $c$ & $S_0$ \\
$1/2$   & 1.02288 $\pm$ 0.00017 & -0.26705 $\pm$ 0.00026    \\
$3/2$   & 1.13639 $\pm$ 0.01468 & -0.01619 $\pm$ 0.02340    \\
$5/2$   & 1.15087 $\pm$ 0.01912 & -0.01018 $\pm$ 0.03059    \\[4pt] \hline
\multicolumn{3}{c}{$K=5.0$\Tstrut}                        \\
  $l_{\mathrm{max}}$   & $c$ & $S_0$ \\
$1/2$   & 1.02244 $\pm$ 0.00019 & -0.26632 $\pm$ 0.00030    \\
$3/2$   & 1.20910 $\pm$ 0.01398 & -0.03521 $\pm$ 0.02237    \\
$5/2$   & 1.24826 $\pm$ 0.02317 & -0.04904 $\pm$ 0.03780    \\[4pt] \hline
\multicolumn{3}{c}{$K=6.0$\Tstrut}                        \\
  $l_{\mathrm{max}}$   & $c$ & $S_0$ \\
$1/2$   & 1.02187 $\pm$ 0.00017 & -0.26544 $\pm$ 0.00027    \\
$3/2$   & 1.26305 $\pm$ 0.01474 & -0.06317 $\pm$ 0.02372    \\
$5/2$   & 1.32593 $\pm$ 0.02497 & -0.10043 $\pm$ 0.04088    \\[4pt] 
\bottomrule 
\end{tabular}
\end{threeparttable}
\caption{Linear fitting results of the bipartite entanglement entropy data.
For each $K$ and $l_{\mathrm{max}}$, we list the estimation and the error for the central charge $c$ (from the slope) and the intercept $S_0$. 
The errors originate from both the uncertainty in the extrapolation of the original data to infinite bond dimension and the uncertainty in the linear fitting.
}
\label{tab:entanglement-fits}
\end{table}

\section{Summary and outlook} \label{sec:summary}

In summary, we have numerically studied the (1+1)-dimensional $O(3)$ nonlinear $\sigma$-model with $\theta=\pi$ term using our recently developed cMPO method.
We work in the Hamiltonian formulation---the modified quantum rotor model decorated with magnetic monopoles, and derive its matrix presentation in the monopole harmonics basis.
From this matrix representation, we obtain the cMPO representation of the modified quantum rotor model, and study its finite-temperature properties with the cMPO method. 
We calculate the free energy density of the system and the bipartite entanglement entropy of the boundary cMPS, and compare their scaling with the predictions of CFT, from which we confirm the massless nature of the system and obtain the central charge as a function of the coupling constant. 
The Hamiltonian formalism and the corresponding matrix representation for NLSM with $\theta=\pi$ also enables further studies related to this model, such as the effect of nonzero chemical potential and the real-time dynamics.  

%By working with the Hamiltonian formulation and the cMPO formalism, only the spatial direction is discretized as the imaginary time direction of the model is automatically continuous.
%The discretization effect can be remedied by lowering the temperature. 

For tensor network simulations, we have to truncate the monopole harmonics basis and only consider the states with relatively small angular momentum.
This truncation scheme works well near the strong coupling limit since the high angular momentum sectors are suppressed, and the result can easily get converged with the maximal angular momentum $l_{\mathrm{max}}$.
However, as the system goes toward the weak coupling limit, the angular momentum is no longer suppressed, and the monopole harmonics basis becomes less effective and moderate values of $l_{\mathrm{max}}$ often do not lead to a converged result.
An intermediate improvement based on our formalism is to implement the non-Abelian symmetry in tensor network simulations~\cite{singh-tensor-2012,weichselbaum-nonabelian-2012,schmoll-programming-2020} so that one can push the calculation to larger $l_{\mathrm{max}}$. 
For future investigations, it is also important to look for a better basis near the weak coupling limit in order to truncate the local physical Hilbert space in a more effective way.

Another interesting direction is to consider the possibility of a Hamiltonian formulation for the NLSM with $\theta \neq 0$ or $\pi$.
The current formulation relies on the monopole harmonics basis, which is tied to the quantization of the magnetic monopole charge and does not seem to have a straightforward generalization to $\theta \in (0,\pi)$ cases.
Undoubtedly, the capability to efficiently simulate the (1+1)-dimensional NLSM with a full range of $\theta$ will open a broad way to better understand this paradigmatic field theory.

\section*{Acknowledgment} 
We thank Hai-Jun Liao for helpful discussions. 
W. T. and X. C. X. are supported by the National Basic Research Program of China (Grants No. 2015CB921102, and No. 2017YFA0303301). 
L. W. is supported by the National Natural Science Foundation of China under Grant No.~11774398 and the Strategic Priority Research Program of the Chinese Academy of Sciences under Grant No. XDB30000000. 
H.-H. T. is supported by the Deutsche Forchungsgemeinschaft (DFG) through project A06 of SFB 1143 (project-id 247310070).

\appendix

\section{Matrix representation for the modified quantum rotor models in the angular momentum eigenbasis} \label{app:matrix-representation}

In this Appendix, we derive the matrix representation for the modified quantum rotor model in Eq.~\eqref{eq:qrm-thetapi}.
Apparently, the kinetic term is diagonal in the angular momentum basis 
\begin{equation}
  \<q,l,m | \frac{(\hat{\boldL}')^2}{2K} | q,l,m \> = \frac{l^2}{2K},
\end{equation}
and we only need to focus on the coupling term between neighboring rotors. 
The $\hat{\boldn}_i \cdot \hat{\boldn}_j$ term can be represented as
\begin{equation}
  \hat{\boldn}_i \cdot \hat{\boldn}_j = \frac{1}{2} (\hat{n}_i^+ \hat{n}_j^-  +  \hat{n}_i^- \hat{n}_j^+) + \hat{n}_i^z \hat{n}_j^z, 
\end{equation}
where $\hat{n}^{\pm} = (\hat{n}^x \pm \hat{n}^y) / \sqrt{2}$. 
The matrix representation of $\hat{n}^{\pm}$ and $\hat{n}^z$ can be evaluated by noticing their relations with the spherical tensor operators~\cite{bruckmann-O3-2019}. More specifically, we have 
\begin{align}
  \hat{n}^z &= \sqrt{\frac{4\pi}{3}} \hat{Y}_{1,0}, \\
  \frac{1}{\sqrt{2}}\hat{n}^+ &= - \sqrt{\frac{4\pi}{3}} \hat{Y}_{1,1}, \\
  \frac{1}{\sqrt{2}}\hat{n}^- &= \sqrt{\frac{4\pi}{3}} \hat{Y}_{1,-1}, 
\end{align} 
where $\hat{Y}_{1, M}$ ($M=0, \pm 1$) is the spherical tensor operator of rank 1. The matrix elements of these operators can be computed as integrals over monopole harmonics~\cite{wu-some-1977} 
\begin{align}
  &\phantom{==} \langle q, l_1, m_1 | \hat{Y}_{1, M} | q, l_2, m_2 \rangle \nonumber \\
 &= (-1)^{q+m_1}\int \mathd \boldn \, Y_{-q, l_1, -m_1}(\boldn) Y_{1, M}(\boldn) Y_{q, l_2, m_2}(\boldn) \nonumber \\
 &= (-1)^{l_1+l_2+1} (-1)^{q + m_1} \sqrt{\frac{3(2l_1+1)(2l_2+2)}{4\pi}}  \nonumber \\
 &\phantom{=} \times
 \begin{pmatrix}
   l_1 & 1 & l_2 \\
   -m_1 & M & m_2
   \end{pmatrix}
   \begin{pmatrix}
   l_1 & 1 & l_2 \\
   -1/2 & 0 & 1/2
   \end{pmatrix},
\end{align}
where $\begin{pmatrix}
    j_1 & j_2 & j_3 \\ 
    m_1 & m_2 & m_3 
    \end{pmatrix}$ is the Wigner-3j symbol. 

\section{Proof for the mapping between the quantum rotor model and NLSM} \label{app:proof-mapping}

In this Appendix, we present a proof for the mapping between the quantum rotor model and the NLSM based on the path-integral formalism. 
Our proof covers both $\theta = 0$ and $\theta = \pi$ cases.

\subsection{$\theta=0$ case}

For the $\theta = 0$ case, we write the partition function of the ordinary quantum rotor model in Eq.~\eqref{eq:qrm-theta0} as a path integral
\begin{equation}
  Z = \tr (\mathe^{-\beta \hat{H}}) = \int \left(\prod_{k=0}^{N-1} \mathd n^{(k)} \right) 
  \left[ \prod_{k=0}^{N-1}
  \langle n^{(k+1)} | \mathe^{-\Delta \tau \hat{H}}| n^{(k)} \rangle 
  \right], 
\label{eq:partition-func-o3-topo0}
\end{equation}
where the imaginary time $\tau \in [0,\beta]$ is divided into $N$ intervals, i.e., $\Delta \tau = \beta / N$, and $n^{(k)} \equiv (\boldn^{(k)}_1, \boldn^{(k)}_2, \ldots, \boldn^{(k)}_L)$ represents the rotor configuration at the $k$th time interval. The periodic boundary condition $n^{(N)} = n^{(0)}$ along the imaginary time direction is imposed.
Substituting the explicit form of $\hat{H}$ into \Eq{eq:partition-func-o3-topo0}, we obtain
\begin{align}
  \langle n^{(k+1)} | \mathe^{-\Delta \tau \hat{H}}| n^{(k)} \rangle
  & = \left( 
  \prod_j \langle \boldn_j^{(k+1)} | \mathe^{-\Delta \tau \hat{\boldL}_j^2/2Ka} | \boldn_j^{(k)} \rangle
  \right) \nonumber \\
  & \phantom{=} \times 
  \mathe^{\Delta \tau K / a \sum_{j=1}^L \boldn_j^{(k)} \cdot \boldn_{j+1}^{(k)} }.
  \label{eq:o3-topo0-timeseg}
\end{align}
Within \Eq{eq:o3-topo0-timeseg}, we look at $\langle \boldn_j^{(k+1)} | \mathe^{-\Delta \tau \hat{\boldL}_j^2/2Ka} | \boldn_j^{(k)} \rangle$ first. 
By inserting a complete set of angular momentum basis into it, we get 
\begin{align}
  & \phantom{==} \langle \boldn_j^{(k+1)} | \mathe^{-\Delta \tau \hat{\boldL}_j^2/2Ka} | \boldn_j^{(k)} \rangle \nonumber\\ 
  & = \sum_{l=0}^{\infty} \sum_{m=-l}^{l} \langle \boldn^{(k+1)}_j | l,m\rangle \langle l,m| \boldn_j^{(k)} \rangle \mathe^{-\Delta \tau l(l+1)/2Ka}.
\end{align}
The summation over $m$ can be handled by the addition formula of the spherical harmonics, 
\begin{equation}
  P_l(\cos \gamma_j^{(k)}) = \frac{4\pi}{2l+1} \sum_{m=-l}^{l}\langle \boldn^{(k+1)}_j | l,m\rangle \langle l,m| \boldn_j^{(k)} \rangle, 
\end{equation}
where $P_l$ is the Legendre polynomial, and $\gamma_j^{(k)}$ represents the angle between $\boldn_j^{(k)}$ and $\boldn_j^{(k+1)}$.
In the $\Delta \tau \rightarrow 0$ limit,  $\gamma_j^{(k)}$ becomes infinitesimal, and 
\begin{equation}
  P_l (\cos \gamma_j^{(k)}) \approx 1 - \frac{1}{4} l (l+1) (\gamma_j^{(k)})^2 \approx \exp [\frac{1}{4}l(l+1)(\gamma_j^{(k)})^2].
\end{equation}
Therefore, 
\begin{align}
&  \phantom{==}  \langle \boldn_j^{(k+1)} | \mathe^{-\Delta \tau \hat{\boldL}_j^2/2K} | \boldn_j^{(k)} \rangle \nonumber \\
& = \sum_{l=0}^{\infty} \frac{2l+1}{4\pi} \exp\left[- l(l+1) \left(\frac{\gamma_j^{(k)})^2}{4} + \frac{\Delta \tau}{2Ka} \right) \right] \label{eq:tmp-sum-exp} \\
& \approx \frac{1}{4\pi} \frac{1}{\frac{\Delta \tau}{2Ka} + \frac{(\gamma_j^{(k)})^2}{4}} \label{eq:tmp-sum-exp-after} \\
& \approx \frac{Ka}{2\pi \Delta \tau} \exp\left[-\frac{Ka \Delta \tau}{2} (\partial_{\tau} \boldn_j^{(k)})^2\right],
  \label{eq:o3-topo0-timeseg-single}
\end{align}
where we have introduced $\partial_{\tau} \boldn_j^{(k)} = \gamma_j^{(k)}/\Delta \tau$. From Eq.~\eqref{eq:tmp-sum-exp} to Eq.~\eqref{eq:tmp-sum-exp-after}, the summation over $l$ is carried out by the Euler-MacLaurin formula. 

Next, we evaluate $\exp[\Delta \tau (K/a) \sum_{j=1}^L \boldn_j^{(k)} \cdot \boldn_{j+1}^{(k)}]$ in \Eq{eq:o3-topo0-timeseg}.  
Here we assume that the lattice spacing $a$ is small, and $\boldn_j^{(k)} \cdot \boldn_{j+1}^{(k)} = \cos(\delta_j^{(k)}) \approx 1 - (1/2) (\delta_j^{(k)})^2$, where $\delta_j^{(k)}$ represents the angle between $\boldn_j^{(k)}$ and $\boldn_{j+1}^{(k)}$.
Along this line, we get
\begin{equation}
\exp\left(\frac{\Delta \tau K}{a} \sum_{j=1}^L \boldn_j^{(k)} \cdot \boldn_{j+1}^{(k)} \right)
= \mathe^{\Delta \tau K L/a} \exp(-\frac{\Delta \tau K a}{2} \sum_{j=1}^L (\partial_x \boldn_j^{(k)})^2),
\label{eq:o3-topo0-spatial-sg}
\end{equation}
where we have introduced $\partial_x \boldn_j^{(k)} = \delta_j^{(k)} / a$.

Combining Eqs.~\eqref{eq:o3-topo0-timeseg}, \eqref{eq:o3-topo0-timeseg-single}, and \eqref{eq:o3-topo0-spatial-sg}, we get
\begin{align}
  & \phantom{==} \langle n^{(k+1)} | \mathe^{-\Delta \tau \hat{H}}| n^{(k)} \rangle \nonumber \\ 
  & = \left(\frac{Ka \mathe^{\Delta \tau K/a}}{2\pi \Delta \tau}\right)^L 
  \exp\left\{-\frac{K \Delta \tau a}{2} \sum_{j,k} \left[(\partial_{\tau} \boldn_j^{(k)})^2 + (\partial_x \boldn_j^{(k)})^2 \right] \right\}.
\end{align}
Substituting this equation into \Eq{eq:partition-func-o3-topo0}, and taking the limit $a, \Delta \tau \rightarrow 0$, we find
\begin{equation}
  Z = \int \mathD n(x, \tau) \, \exp\left[ -\int_0^L \mathd x \int_0^{\beta} \mathd \tau \, \mathcal{L}(x, \tau) \right],
\end{equation}
where $\mathD n(x, \tau) \equiv \lim_{a\rightarrow 0} \lim_{\Delta\tau \rightarrow 0} \prod_{j,k} (Ka\mathe^{\Delta \tau K / a} / 2\pi \Delta \tau)$, and the Lagrangian density $\mathcal{L}$ is given by 
\begin{equation}
  \mathcal{L} = \frac{K}{2} \left[ (\partial_{\tau} \boldn(x, \tau))^2 + (\partial_x \boldn(x, \tau))^2 \right].
\end{equation}

\subsection{$\theta=\pi$ case}

For the NLSM with $\theta=\pi$ term, similarly as Eq.~\eqref{eq:partition-func-o3-topo0}, we write the partition function of \Eq{eq:qrm-thetapi} as  
\begin{equation}
  Z = \tr (\mathe^{-\beta \hat{H}}) = \int \left(\prod_{k=0}^{N-1} \mathd n^{(k)} \right) 
  \left[ \prod_{k=0}^{N-1}
  \langle n^{(k+1)} | \mathe^{-\Delta \tau \hat{H}}| n^{(k)} \rangle 
  \right], 
\label{eq:partition-func-o3-topopi}
\end{equation}
where $n^{(k)} \equiv (\boldn^{(k)}_1, \boldn^{(k)}_2, \ldots, \boldn^{(k)}_L)$ denotes the $k$th configurations of the quantum rotors, and $n^{(N)} = n^{(0)}$.
At each time slice, we have
\begin{align}
  & \phantom{==} \langle n^{(k+1)} | \mathe^{-\Delta \tau \hat{H}}| n^{(k)} \rangle \nonumber \\
  & = \left( 
  \prod_j \langle \boldn_j^{(k+1)} | \mathe^{-\Delta \tau (\hat{\boldL}'_j)^2/2Ka} | \boldn_j^{(k)} \rangle
  \right)
  \exp\left[\frac{\Delta \tau K}{a} \sum_{j}^L \boldn_j^{(k)} \cdot \boldn_{j+1}^{(k)} \right],
  \label{eq:o3-topopi-timeseg}
\end{align}
which is similar to \Eq{eq:o3-topo0-timeseg} except the modified angular momentum operator and the antiferromagnetic coupling between the neighboring rotors.

In \Eq{eq:o3-topopi-timeseg}, we first focus on the kinetic terms $\langle \boldn_j^{(k+1)} | \mathe^{-\Delta \tau (\hat{\boldL}'_j)^2/2Ka} | \boldn_j^{(k)} \rangle$, which can be evaluated as 
\begin{align}
  &\phantom{==} \langle \boldn_j^{(k+1)} | \mathe^{-\Delta \tau (\hat{\boldL}'_j)^2/2Ka} | \boldn_j^{(k)} \rangle \nonumber \\
  &= \sum_{l=1/2}^\infty \sum_{m=-l}^{l} \langle \boldn_j^{(k+1)} | q,l,m \rangle \langle q,l,m | \boldn_j^{(k)}\rangle \exp\left(-\frac{\Delta \tau l (l+1)}{2Ka}\right),
  \label{eq:o3-topopi-timeseg0} 
\end{align}
where $|q, l, m\rangle$ represents the eigenbasis of the modified angular momentum operator (where $q=1/2$).
Using the addition formula of the monopole harmonics~\cite{wu-dirac-1976,wu-some-1977}, the summation over $m$ can be carried out, and we get 
\begin{align}
  &\phantom{==} \sum_{m=-l}^{l} \langle \boldn_j^{(k+1)} | q,l,m \rangle \langle q,l,m | \boldn_j^{(k)}\rangle  \nonumber \\
  &= \frac{2l+1}{4\sqrt{2}\pi} \sqrt{1+\cos\gamma_j^{(k)}} P_{l-1/2}^{0,1}(\cos\gamma_j^{(k)})) \mathe^{-\mathi \Omega_j^{(k)} / 2},
  \label{eq:o3-topopi-timeseg1}
\end{align}
where $P_n^{\mu, \nu}$ is the Jacobi polynomial, $\gamma_j^{(k)}$ represents the angle between $\boldn_j{(k)}$ and $\boldn_j{(k+1)}$. $\Omega_j^{(k)}$ is the area of the spherical triangular formed by $\boldn_j{(k)}$, $\boldn_j{(k+1)}$, and the north-pole axis on the unit sphere.
As $\gamma_j^{(k)} \ll 1$, we have $\cos\gamma_j^{(k)} \approx 1 - (1/2)(\gamma_j^{(k)})^2$, and \Eq{eq:o3-topopi-timeseg1} can be further simplified as
\begin{align}
  &\phantom{==} \sum_{m=-l}^{l} \langle \boldn_j^{(k+1)} | q,l,m \rangle \langle q,l,m | \boldn_j^{(k)}\rangle \nonumber \\
  &\approx \frac{2l+1}{4\pi} \exp\left[ \frac{(\gamma_j^{(k)})^2}{4} \left(l^2+l-\frac{1}{4}\right) - \frac{\mathi \Omega_j^{(k)}}{2} \right]. 
  \label{eq:o3-topopi-timeseg2}
\end{align}
Substituting \Eq{eq:o3-topopi-timeseg2} into \Eq{eq:o3-topopi-timeseg}, and performing the summation over $l$ with the Euler-MacLaurin formula, we arrive at  
\begin{equation}
  \langle \boldn_j^{(k+1)} | \mathe^{-\Delta \tau \hat{H}} | \boldn_j^{(k+1)} \rangle 
  \approx \frac{Ka}{2\pi \Delta} \exp\left[ -\frac{Ka\Delta \tau}{2} (\partial_{\tau} \boldn_j^{(k)})^2 - \frac{\mathi \Omega_j^{(k)}}{2} \right],
  \label{eq:o3-topopi-timeseg3}
\end{equation}
where have introduced $\partial_{\tau} \boldn_j^{(k)} \equiv \gamma_j^{(k)}/ \Delta \tau$.

Combining \Eq{eq:o3-topopi-timeseg3} with Eqs.~\eqref{eq:partition-func-o3-topopi} and \eqref{eq:o3-topopi-timeseg}, we get 
\begin{align}
  Z = & \int \left( \prod_{j, k} \mathd \boldn_j^{(k)} \right)
       \left( \frac{Ka}{2\pi \Delta} \right)^L \times \nonumber \\ 
      & \exp \left[  \sum_{j,k} 
      \left(
      -\frac{K a \Delta \tau}{2} (\partial_{\tau} \boldn_j^{(k)})^2 
      +\frac{\Delta \tau K}{a} \boldn_j^{(k)} \cdot \boldn_{j+1}^{(k)}
      - \frac{\mathi \Omega_j^{(k)}}{2} 
      \right)
      \right].
      \label{eq:partition-func-o3-topopi1}
\end{align}
The continuum in the imaginary time direction can be straightforwardly taken. 
To take the continuum limit in the spatial direction, we split the $\boldn_j^{(k)}$ into the slowly varying part $\boldm_j^{(k)}$ and the rapidly varying part $\boldl_j^{(k)}$,
\begin{equation}
  \boldn_j^{(k)} = (-1)^j \boldm_j^{(k)} + a \boldl_j^{(k)},
  \label{eq:rotor-seperate}
\end{equation} 
where $a$ is the lattice spacing.
Substituting \Eq{eq:rotor-seperate} into terms in \Eq{eq:partition-func-o3-topopi1}, we get 
\begin{align}
   (\partial_{\tau} \boldn_j^{(k)})^2 &= (\partial_{\tau} \boldm_j^{(k)})^2 + 2 a (-1)^j \partial_{\tau} \boldm_m^{(k)} \cdot \partial_\tau \boldl_j^{(k)}, \label{eq:slow-fast-sep1}\\
   \boldn_j^{(k)} \cdot \boldn_{j+1}^{(k)} &= - \boldm_j^{(k)} \cdot \boldm_{j+1}^{(k)} 
   + (-1)^j a (\boldm_j^{(k)}\cdot\boldl_{j+1}^{(k)} - \boldm_{j+1}^{(k)}\cdot\boldl_{j}^{(k)}) \nonumber \\
   &= - 1 + \frac{a^2}{2} (\partial_x \boldm_j^{(k)})^2 + (-1)^j a (\boldm_j^{(k)}\cdot\boldl_{j+1}^{(k)} - \boldm_{j+1}^{(k)}\cdot\boldl_{j}^{(k)}),
   \label{eq:slow-fast-sep2}
\end{align}
where we have introduced $\partial_x \boldm_j^{(k)} = (\boldm_j^{(k)} - \boldm_{j-1}^{(k)}) / a$, and the higher-order of $a$ are neglected.
The surface term $\Omega_j^{(k)}$ requires a more careful treatment. 
Since the neighboring rotors tend to have an antiparallel alignment, we separate the summation of surface terms in pairs. We look at each of these pairs
\begin{equation}
  \Delta S (2r) = \sum_{k} (\Omega_{2r}^{(k)} +\Omega_{2r-1}^{(k)}),
\end{equation}
which corresponds to the surface area of the ribbon formed by the trajectories of $\boldn_{2r}^{(k)}$ and $-\boldn_{2r-1}^{(k)}$ on the unit sphere. 
We calculate $\Delta S(2r)$ as 
\begin{align}
  \Delta S(2r) &= \Delta \tau \sum_{k=0}^{N-1} \left(\boldn_{2r}^{(k)} + \boldn_{2r-1}^{(k)} \right) \cdot \left(\boldn_{2r}^{(k)} \times \partial_\tau \boldn_{2r}^{(k)} \right)\nonumber \\
  & = \Delta \tau a \sum_{k=0}^{N-1} \left(\partial_x \boldm_{2r}^{(l)} + \boldl_{2r}^{(k)} + \boldl_{2r-1}^{(k)} \right) \cdot \left( \boldm_{2r}^{(k)} \times \partial_\tau \boldm_{2r}^{(k)} \right).
  \label{eq:slow-fast-sep3}
\end{align}
Substituting Eqs.~\eqref{eq:slow-fast-sep1}, \eqref{eq:slow-fast-sep2}, \eqref{eq:slow-fast-sep3} into \Eq{eq:partition-func-o3-topopi1}, and integrating out the fast fields $\boldl_j^{(k)}$, we get
\begin{align}
  Z = & \int \left( \prod_{j, k} \mathd \boldm_j^{(k)} \right)
        \exp \left[  -\frac{K a \Delta \tau}{2} \sum_{j,k} 
        \left(
        (\partial_{\tau} \boldm_j^{(k)})^2 
        + (\partial_{x} \boldm_j^{(k)})^2 
        \right)
      \right] \nonumber \\ 
      & \times \exp \left[- \frac{\mathi a \Delta \tau}{4} \sum_{j, k} \boldm_j^{(k)} \cdot \left( \partial_\tau \boldm_j^{(k)} \times \partial_x \boldm_j^{(k)} \right)  \right].
\end{align} 
Finally, taking the continuum limit, we arrive at 
\begin{equation}
  Z = \int \mathD \boldm(x, \tau) \, \exp(-\int_0^L \mathd x \int_{0}^{\beta} \mathd \tau \, \mathcal{L} (x, \tau)),
\end{equation}
where the Lagrangian density is given by
\begin{align}
  \mathcal{L} = & \frac{K}{2} \int \mathd x \mathd \tau \, 
   \left[ 
    (\partial_x \boldm(x, \tau))^2 + (\partial_\tau \boldm(x, \tau))^2
  \right] \nonumber \\
  & + \frac{\mathi \theta}{4\pi} \int \mathd x \mathd \tau \,
  \boldm(x, \tau) \cdot (\partial_\tau \boldm(x, \tau) \times \partial_x \boldm(x, \tau))
\end{align}
with $\theta = \pi$.

\section{Numerical details} \label{app:details}

\subsection{Determination of degeneracy sectors in \boldmath$U(1)$-symmetric boundary cMPS} \label{app:u1}

In the cMPO approach, the boundary cMPS is uniform, and has periodic boundary condition (PBC).
For the $U(1)$-symmetric boundary cMPS, one has to determine the degeneracy sectors on the vertical bonds, which further determines the block structures in the cMPS local tensors.
However, unlike the open-boundary MPS simulations, it is difficult to determine an optimal choice for these degeneracy sectors. 
%One possible choice is to test all possible choices and compare the result of the free energy densities. 
In our simulation, we dynamically determine the degeneracy sectors during a power method process. 

\begin{figure}[!htb]
  \centering
  \resizebox{0.9\columnwidth}{!}{\includegraphics{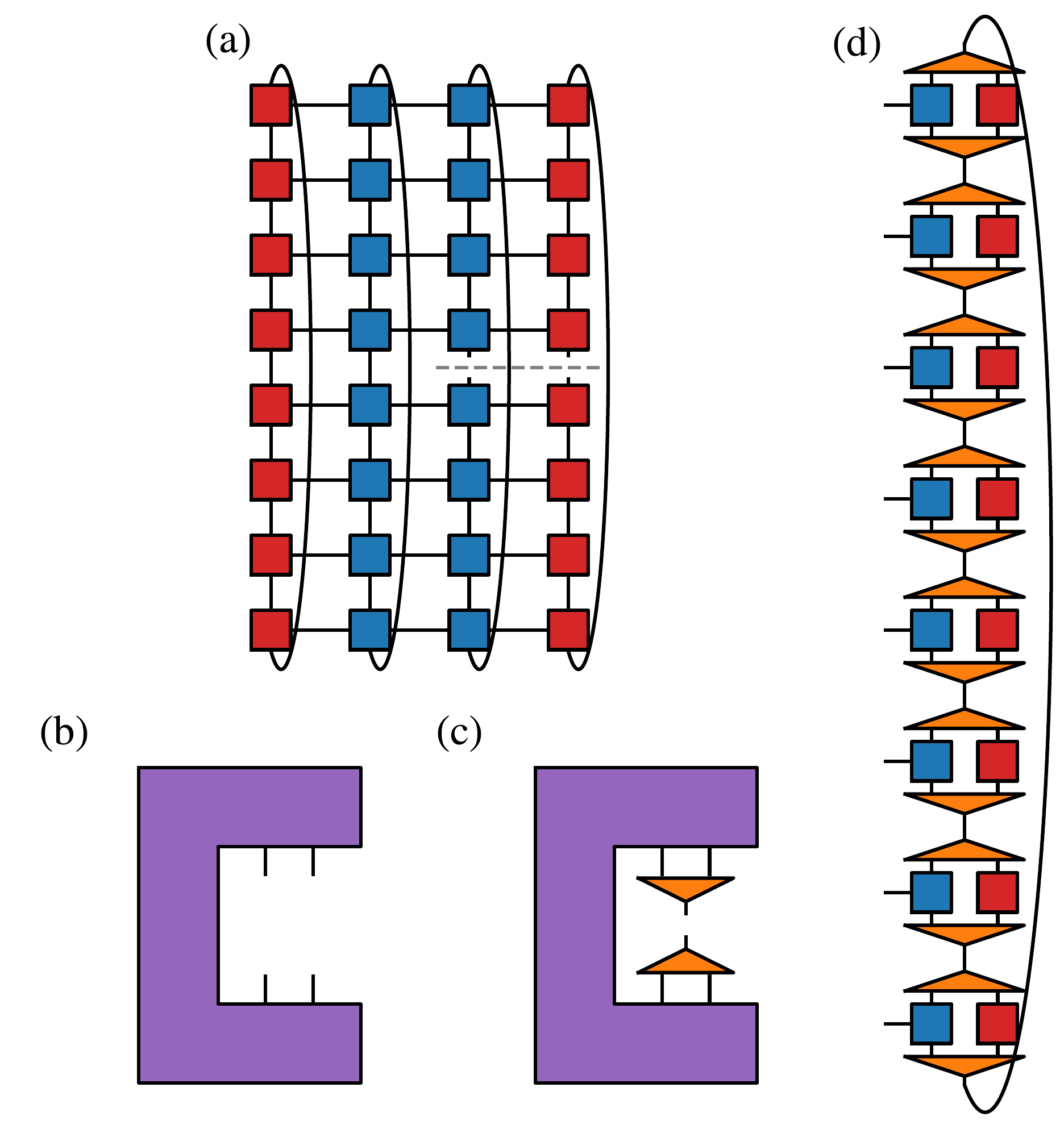}}
  \caption{The initialization of the variational compression by inserting isometries into the cMPS.  (a) The tensor network representation of the reduced density matrix. The red and blue tensors represent the cMPS and cMPO tensors respectively. (b) The reduced density matrix. (c) The truncation of the reduced density matrix which gives the isometry. (d) Truncation by inserting the isometries into the cMPS.}
  \label{fig:isometry-illustration}
\end{figure}

As is discussed in Sec.~\ref{sec:cmpo-formulation-qrm}, we initialize the variational optimization of the boundary cMPS by a few steps of power method process.
To start with, we construct an initial cMPS by using the cMPO tensors at the boundary sites. 
Using the right boundary cMPS as an example, the local tensor $T_{\psi}^{(0)}$ of the initial cMPS $|\psi^{(0)} \rangle$ can be obtained from the local tensor $T$ [see Eqs.~\eqref{eq:cmpo-general} and \eqref{eq:cmps}],
\begin{equation}
  T =
\left( \begin{array}{cc}
  I + \epsilon \sfQ & \sqrt{\epsilon} \sfR \\
  \sqrt{\epsilon} \sfL & \sfP 
  \end{array} \right) 
  \rightarrow 
  T_{\psi}^{(0)} = 
\left( \begin{array}{c}
  I + \epsilon \sfQ  \\
  \sqrt{\epsilon} \sfL 
  \end{array} \right).
\end{equation}
The $U(1)$ degeneracy sectors for $|\psi^{(0)}\rangle$ are simply derived from those for the cMPO. 
Next, we repeatedly act the cMPO $\mathbb{T}$ on the cMPS, and compress the cMPS variationally if the bond dimension exceeds the target bond dimension $\chi$.
By considering the truncation scheme illustrated in Fig.~\ref{fig:isometry-illustration}, we initialize the variational compression process and determines the degeneracy sectors.
Suppose we are at the $n$th power method step, and we act the cMPO $\mathbb{T}$ on the cMPS $|\psi^{(n)}\rangle$. 
We construct a reduced density matrix $\rho^{(n)}$ by cutting one horizontal bond of $\mathbb{T}|\psi^{(n)} \rangle$ in the tensor network corresponding to $\langle \psi^{(n)}| \mathbb{T}^\dagger \mathbb{T} |\psi^{(n)}\rangle$, as shown in Fig.~\ref{fig:isometry-illustration} (a) and (b).
The reduced density matrix $\rho^{(n)}$ also has a block-diagonalized structure due to the $U(1)$ symmetry, and it shares the same degeneracy structure with $\mathbb{T}|\psi^{(n)}\rangle$. 
After diagonalizing $\rho^{(n)}$, by keeping $\chi$ eigenvalues with the largest abstract values, we can obtain an isometry and insert it into $\mathbb{T}|\psi^{(n)}\rangle$, which can be used as the starting point for the variational compression [see Fig.~\ref{fig:isometry-illustration} (c) (d)]. 
The degeneracy sectors for $\psi^{(n+1)}$ can be automatically determined during this process. 

Finally, after a few power steps described above, we can further optimize the boundary cMPS by variationally minimizing the free energy globally.

\subsection{Extrapolation to the infinite bond dimension} \label{app:extrapolation}

For each set of parameters, the bond dimension for the boundary cMPS ranges among $\chi = 12, 18, 24, 30$.
We eliminate the bias brought by the finite $\chi$ and extrapolate the data to the infinite bond dimension. 

\begin{figure}[!htb]
  \centering
  \resizebox{\columnwidth}{!}{\includegraphics{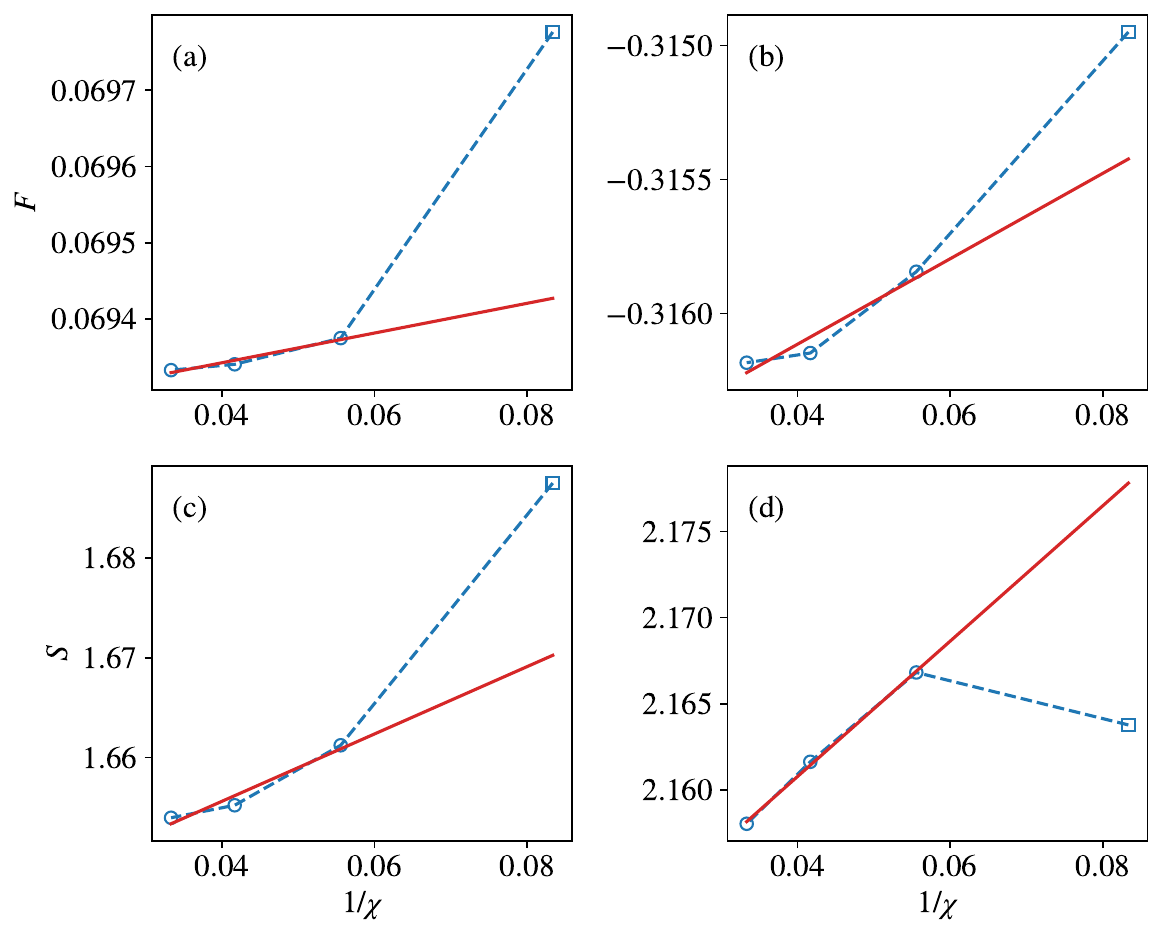}}
  \caption{Examples for the linear extrapolation with respect to $1/\chi$ using the data with the three largest $\chi$'s for $K=2, \lmax=5/2, K\beta=200$ [Figs.~(a), (c)] and $K=5, \lmax=5/2, K\beta=200$ [Figs.~(b), (d)].
  Figures (a), (b) and figures (c), (d) respectively show the linear extrapolation of the free energy and bipartite entanglement entropy.
  In each figure, the circle-shaped dots represent the data points used for the linear regression, and the square-shaped dots represent the unused data. 
  The red solid lines denote the linear fitting of the data.} 
  \label{fig:fit-chi}
\end{figure}

Our extrapolation procedure follows that given in Ref.~\cite{bruckmann-O3-2019}.
First, we perform a linear fitting with respect to $1/\chi$ for the data corresponding to the largest three bond dimensions $\chi = 18, 24, 30$ (see Fig.~\ref{fig:fit-chi}).
The estimation for the extrapolation to infinite bond dimension is obtained by taking the average of the linear extrapolation result and the result corresponding to the largest bond dimension $\chi=30$.
The error for this estimation is given by half of the difference between the linear extrapolation result and the $\chi=30$ result.  

\clearpage

\bibliography{o3nlsmpi}

\bibliographystyle{apsrev4-2}

\end{document}